\documentclass{aa}  
\usepackage{natbib}
\usepackage{graphicx}
\usepackage{txfonts}
\begin{document}
   \title{A new look at a polar crown cavity as observed by SDO/AIA}
   \subtitle{Structure and dynamics}

   \author{S. R\'egnier
	  \and
	  R.~W. Walsh
          \and
          C.~E. Alexander
          }

   \offprints{S. R\'egnier}

   \institute{Jeremiah Horrocks Institute, University of Central Lancashire,
   Preston, Lancashire, PR1 2HE, UK\\
              \email{SRegnier@uclan.ac.uk}
             }

   \date{Received ; accepted }

 
  \abstract
   {The Solar Dynamics Observatory (SDO) was launched in February 2010 and
   is now providing an unprecedented view of the solar activity at high spatial
   resolution and high cadence covering a broad range of temperature layers of
   the atmosphere.}
   {We aim at defining
   the structure of a polar crown cavity and describing its evolution during the
   erupting process. }
   {We use high cadence time series of SDO/AIA observations at 304\AA~(50000 K)
   and 171\AA~(0.6 MK) to determine the structure of the polar crown cavity and
   its associated plasma as well as the evolution of the cavity during the
   different phases of the eruption. }
   {We observe coronal plasma shaped by magnetic field lines with a negative
   curvature (U-shape) sitting at the bottom of a cavity. The cavity is located
   just above the polar crown filament material. We thus observe the inner part
   of the cavity above the filament as depicted in the classical three part
   Coronal Mass Ejection (CME) model composed of a filament, a cavity and a CME
   front. The filament (in this case a polar crown filament) is part of the
   cavity and makes a continuous structuring from the filament to the CME front
   depicted by concentric ellipses (in a 2D cartoon).}
   {We propose to define a polar crown cavity as a density depletion
   sitting above denser polar crown filament plasma drained down the cavity due
   to gravity. As part of the polar crown filament, plasma at different
   temperatures (ranging from 50000K to 0.6 MK) is observed at the same location
   on the cavity dips and sustained by a competition between the gravity and the
   curvature of magnetic field lines. The eruption of the polar crown cavity as
   a solid body can be decomposed into two phases: a slow rise at a speed of 0.6
   km$\cdot$s$^{-1}$, and an acceleration phase at a mean speed of 25
   km$\cdot$s$^{-1}$.}

   \keywords{Sun: corona -- Sun: cavity -- Sun: eruption -- Sun: CME
               }

   \maketitle

\section{Introduction}



Launched in February 2010, the Solar Dynamics Observatory (SDO) is the first NASA
Living With a Star mission. SDO has on board three different instruments
dedicated to study the magnetic and plasma evolution of the solar corona, its
associated eruptive events and their consequences on the Earth. Here we analyse
observations from the Atmospheric Imager Assembly (AIA) with high time cadence
($\sim$ 12s) and  spatial resolution ($\sim$ 1\arcsec) which provide much  more
detail of the corona on the limb than other comparable instruments such as
SOHO/EIT and STEREO/SECCHI/EUVI. Especially it has been shown that SDO/AIA has
the sensitivity to observe far more off-limb structures than never before
\citep{lem11}. Here we report on one of these structures: a cavity
observed on 13 June 2010 associated with an erupting polar crown
filament/prominence leading to a propagating Coronal Mass Ejection (CME) at
a speed of about 300-350 km$\cdot$s$^{-1}$ (as reported in the SOHO/LASCO CME
catalog,  http://cdaw.gsfc.nasa.gov/CME\_list/).

Typically CMEs are observed in white light coronograph
with a three part structure: the bright core related to the filament material,
the cavity surrounding the core and the bright front of the CME marking the
transition between the CME and the ambient corona \citep[see e.g.,][]{ill86}. In
a recent paper by \citet{gib10}, the authors studied in detail the structure,
shape and  evolution (mostly due to the solar rotation) of a stable cavity
observed by SOHO/EIT in the wavelength filter at 195\AA. They found that, based
on a forward modelling approach, the cavity is darker than the surrounding due
to the depletion in density by a factor of about 2. It is clear in their study
that the prominence cavity analysed corresponds to the classical cavity as
defined in the three-part CME model mentioned above. 

The formation and instability of cavities associated with a CME have been
extensively modelled \citep[see reviews by][]{lin02,for06}. The structure often
contains a twisted flux tube able to accumulate magnetic energy and mass within
it which thus can be destabilised by catastrophic mechanisms and/or external
triggers. 
Despite several studies of the thermal structures of cavities especially
from white-light coronographs, eclipse observations, EUV and soft  X-ray
imaging  \citep[e.g.,][]{hud99a,hud00,gib06,hab10}, there is to our knowledge no
observational evidence of a long time series and high cadence obsevations of
cavity at different temperatures as provided by SDO/AIA data able to (i) clearly
demonstrate the thermal structure of both the prominence and cavity material,
and (ii) describe how the plasma of a polar-crown filament evolves before and
during the eruption.

Thus, the observations described here focus for the first time
on the dynamics of the inner part of the cavity above the polar crown
filament/prominence material and its evolution during the eruptive phase.     

We first describe the multithermal structure of the polar crown filament/cavity
(Section~\ref{sec:aia_obs}) and then the evolution of the plasma during the 
eruption in two different wavelengths (Section~\ref{sec:cavity}). In
Section~\ref{sec:disc}, we propose a definition for a polar crown cavity
and discuss the implications of our study on eruptive filament models.

\section{Multithermal observations of the cavity}
\label{sec:aia_obs}

The event was observed on 13 June 2010 between 00:00 and 12:00 UT
on the North-West limb. We focus only on the data provided by the SDO/AIA
instrument using the full spatial resolution and a reduced time cadence of 3
minutes (instead of the nominal 12s). The SDO/AIA data are processed at
level 1 (test series) which includes removal of bad pixels and spikes. The time
series were corrected for pointing and jitter effects. The image
calibration corresponds to a first approximation but does not influence our
study. 

\begin{figure}
\includegraphics[width=0.496\linewidth]{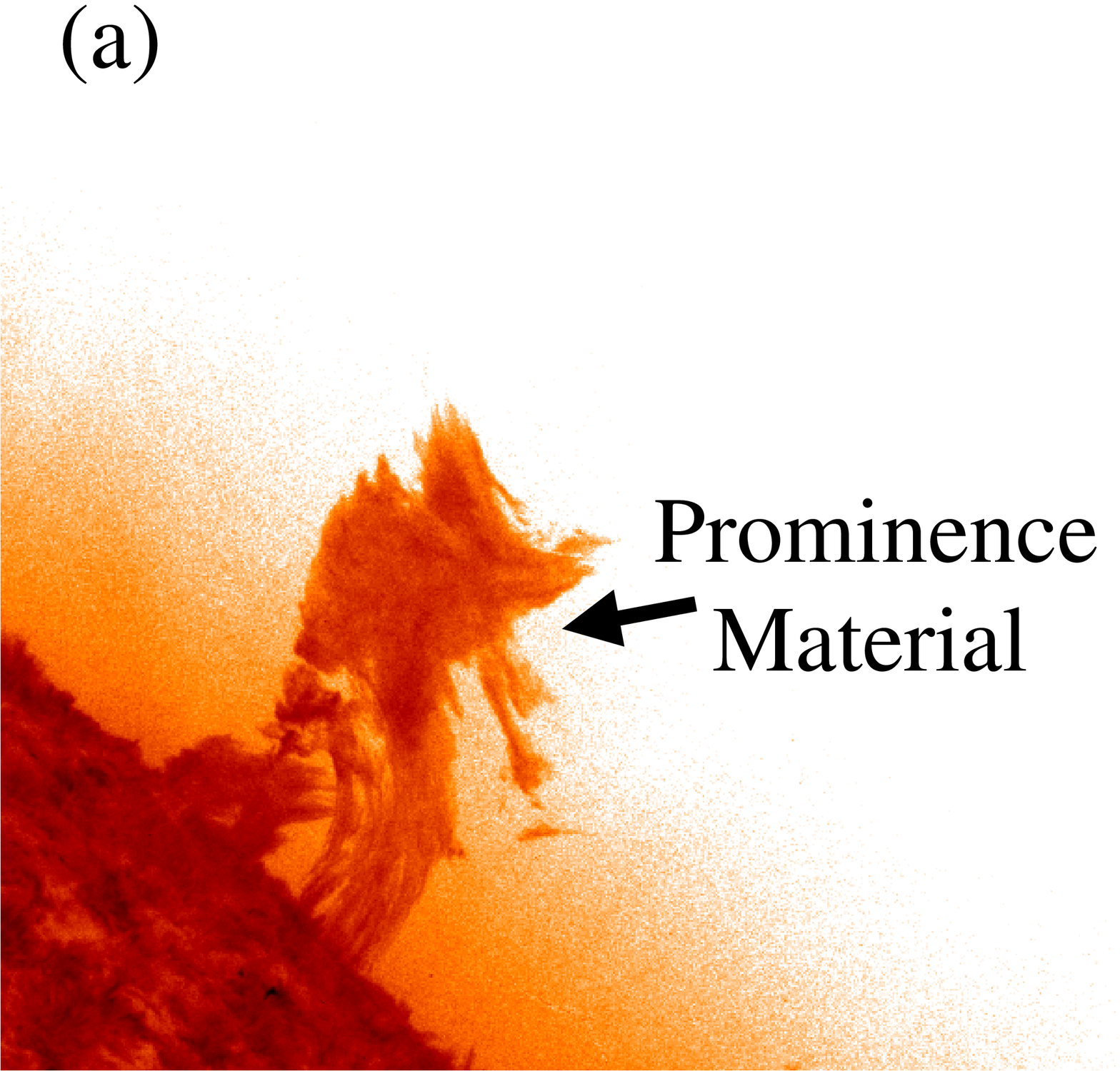}
\includegraphics[width=0.496\linewidth]{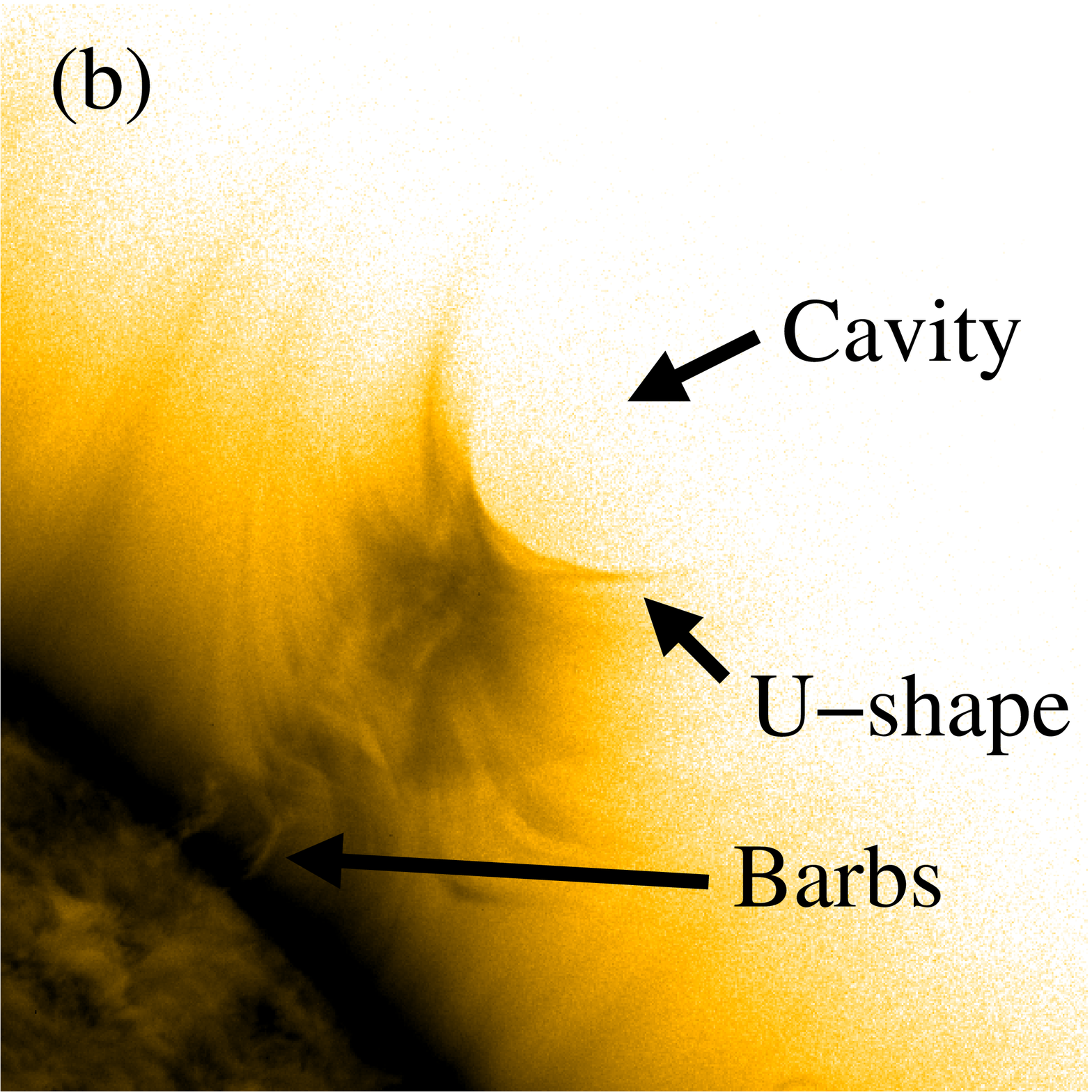}
\includegraphics[width=0.496\linewidth]{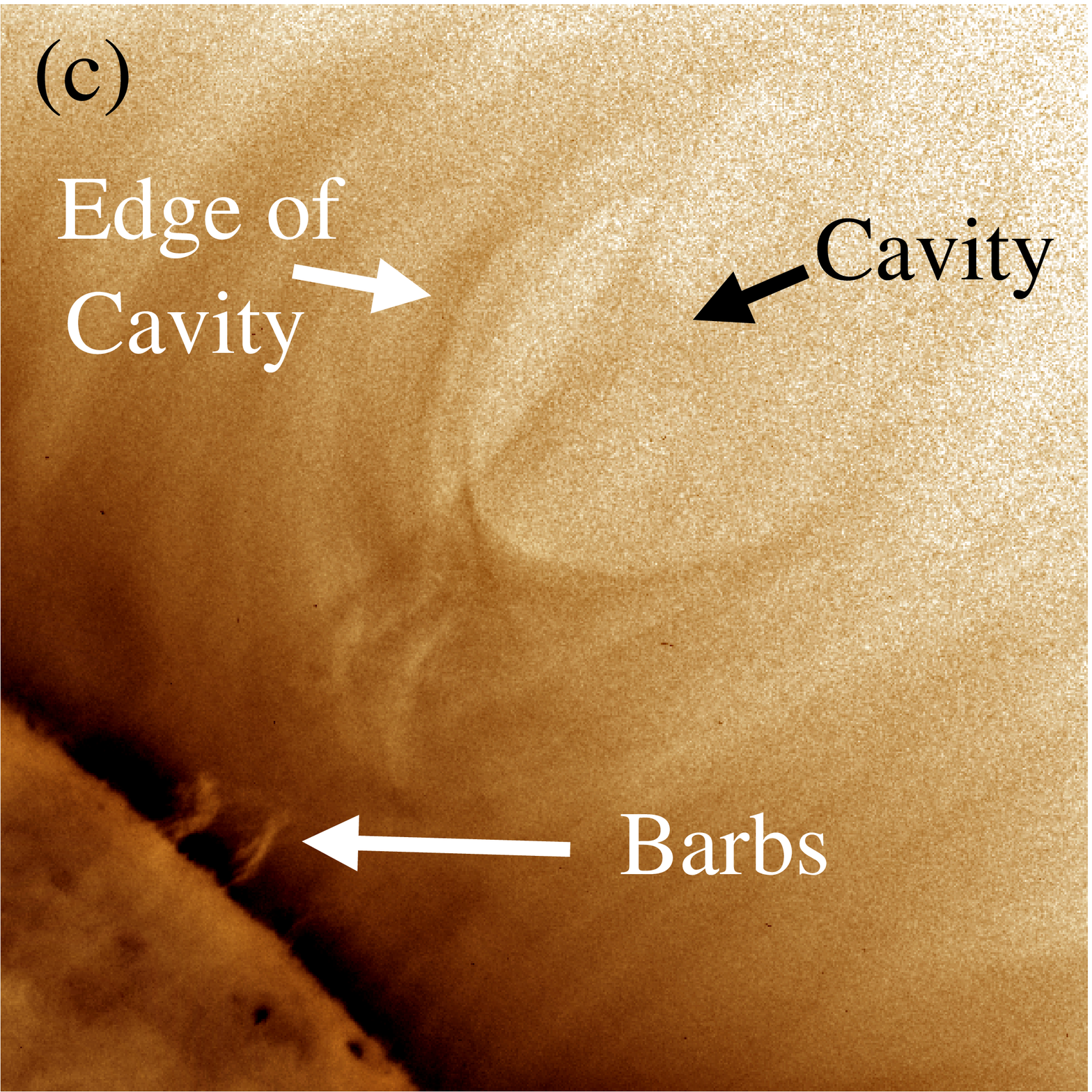}
\includegraphics[width=0.496\linewidth]{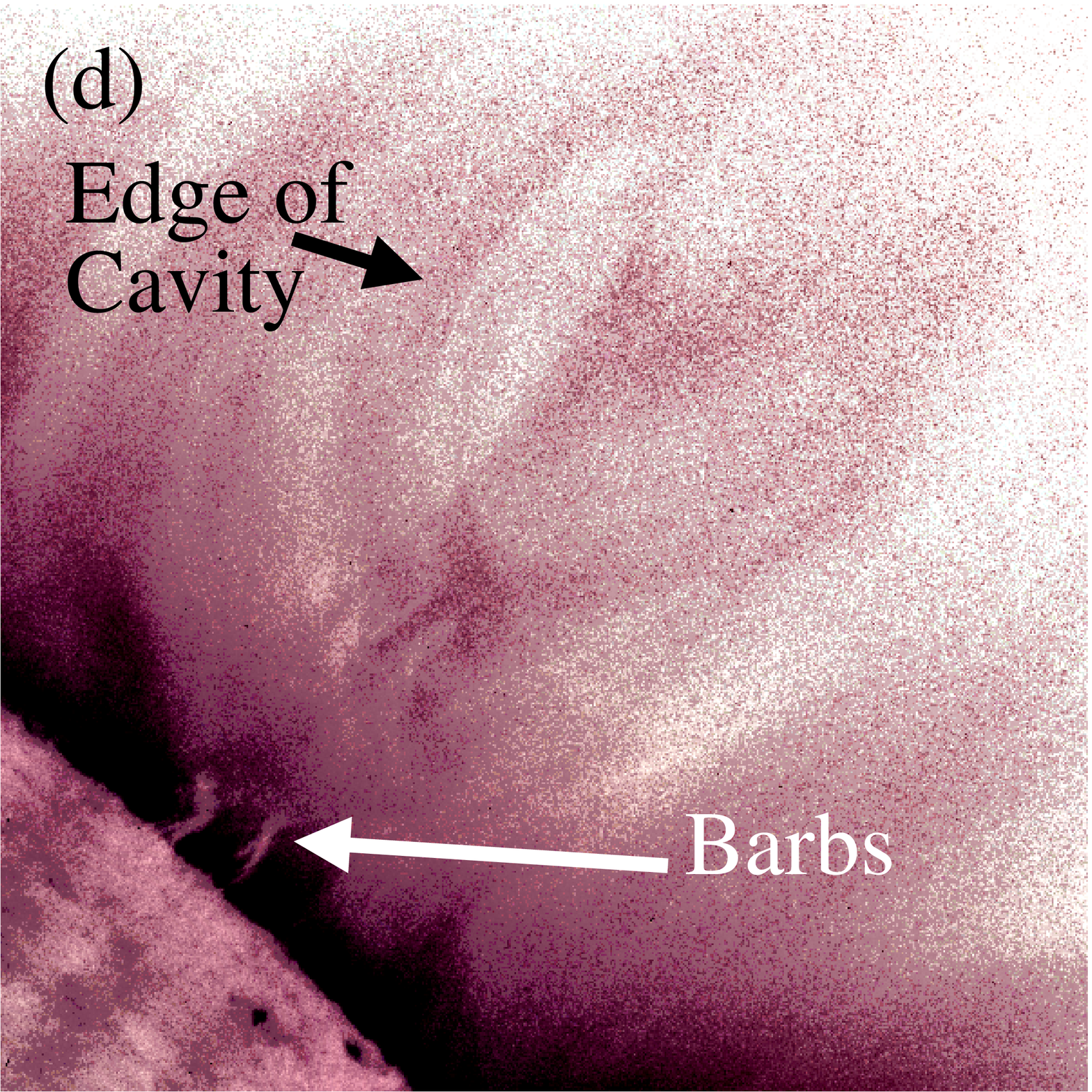}
\caption{Off-limb close-up on the cavity structure as observed by SDO/AIA at
304\AA~(left) and at 171\AA~(right) at 03:24:12 UT (negative images).}
\label{fig:closeup}
\end{figure}

An overview of the structures observed by SDO/AIA is given in
Fig.~\ref{fig:closeup} at different temperatures: (a) HeII at 304\AA ~at about
50000 K, (b) FeIX at 171\AA~at about 0.6 MK, (c) FeXII at 193\AA~at about 1.6 MK
(with a hot contribution of FeXXIV at 2 MK) and (d) FeXIV at 211\AA~at about 2
MK. The single temperature associated with each channel corresponds to the
main peak of emission in the temperature response functions provided by the SDO/AIA
team \citep{lem11}. The study of the SDO/AIA channel thermal response performed
by \citet{odw10} shows the properties of the different SDO/AIA broadband
channels in different regions of the Sun (active region, quiet Sun and coronal
hole). The prominence and cavity material is supposed to be at or near the
temperatures mentioned above. The observed structures are:
\begin{itemize}
\item[(i)]{cool and hot plasma off the limb which are parts of a polar crown
filament (see Fig.~\ref{fig:closeup}a, b). The plasma is confined in an area
with a characteristic height of 100 Mm and width of 80 Mm (at the start of the
time series);}
\item[(ii)]{a dark cavity seen as a complete ellispe in the 193\AA~and
211\AA~channels (see Fig.~\ref{fig:closeup}c, d);
} 
\item[(iii)]{elongated barbs seen as dark material in the hot channels and
connecting the photosphere/chromosphere regions to the cavity (and/or the bottom
of the cavity) to supply or evacuate mass from the filament. We will not discuss
the dynamics of the barbs and their implications in the eruption process in this
letter.}
\end{itemize}

\begin{figure*}[!ht]
\includegraphics[width=0.5\linewidth]{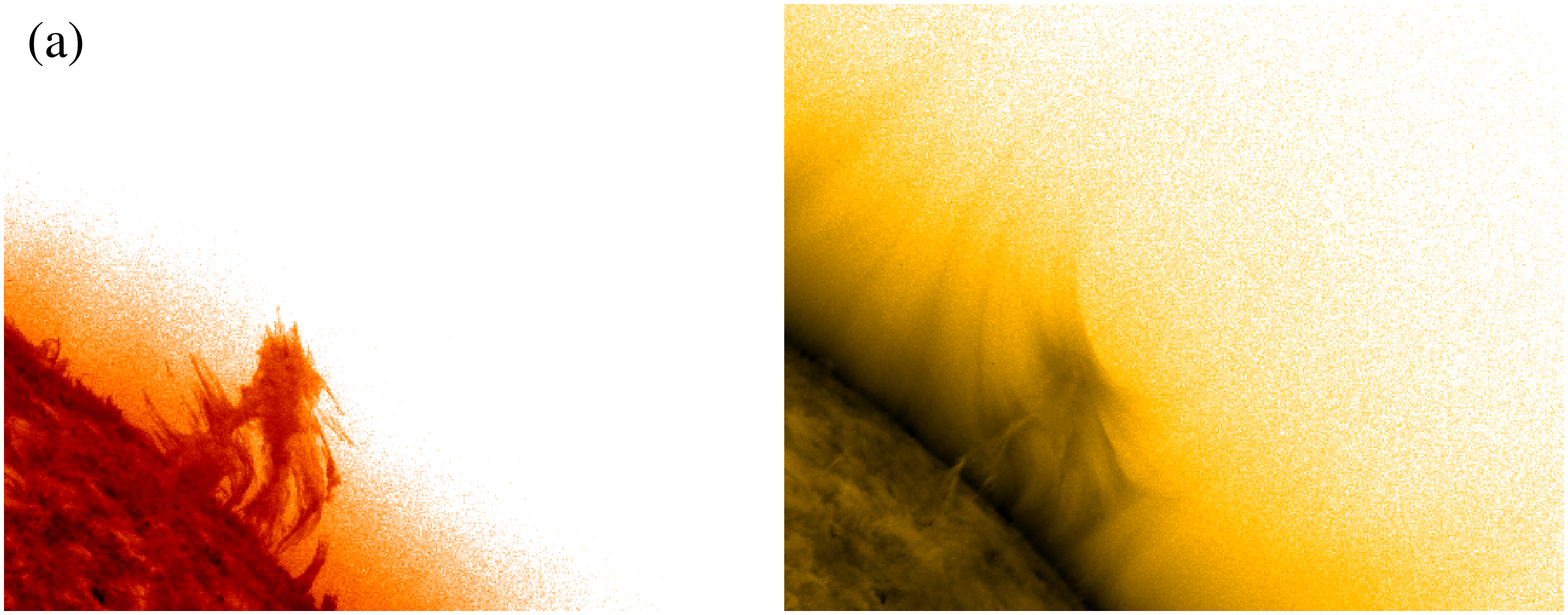}
\includegraphics[width=.5\linewidth]{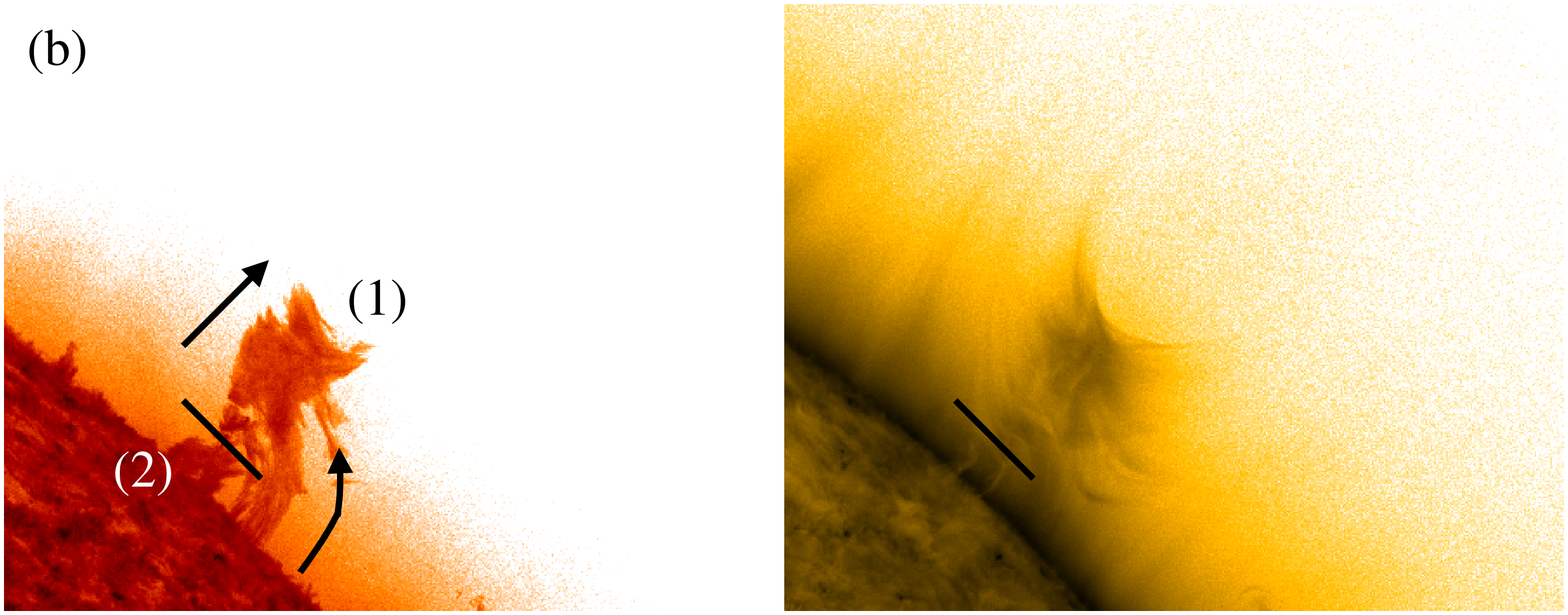}
\includegraphics[width=.5\linewidth]{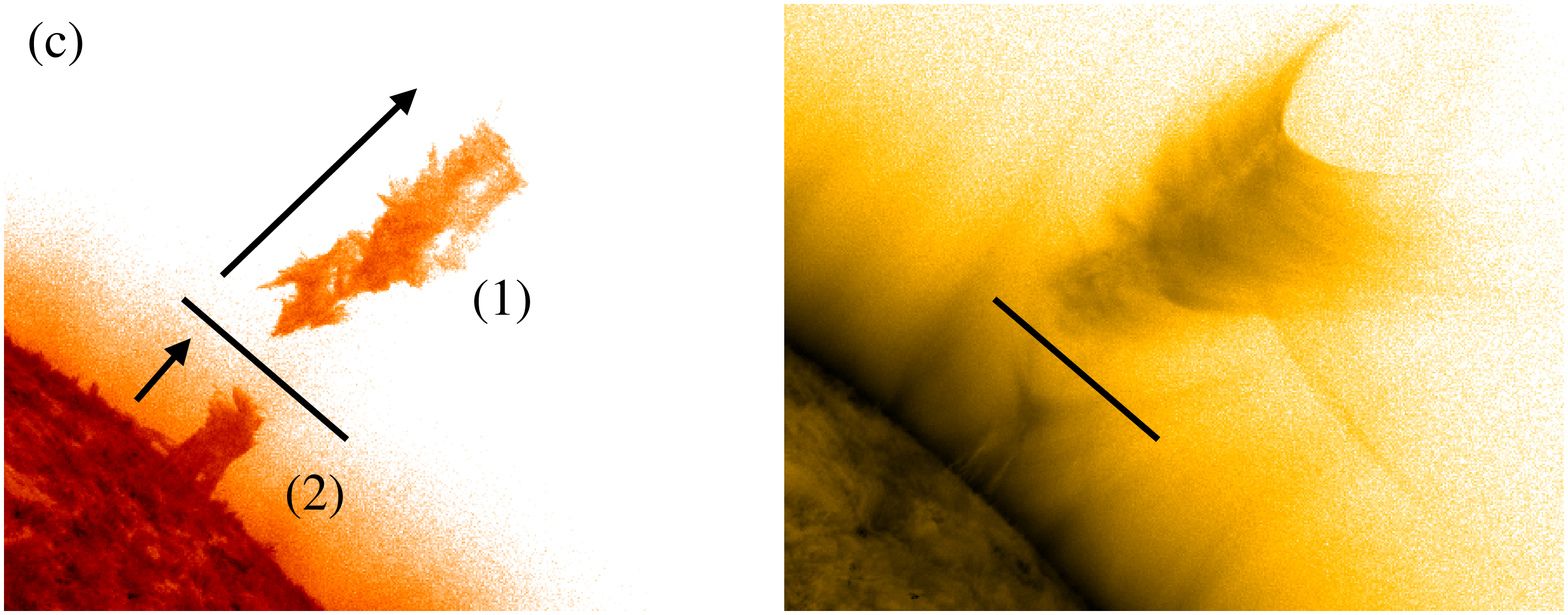}
\includegraphics[width=.5\linewidth]{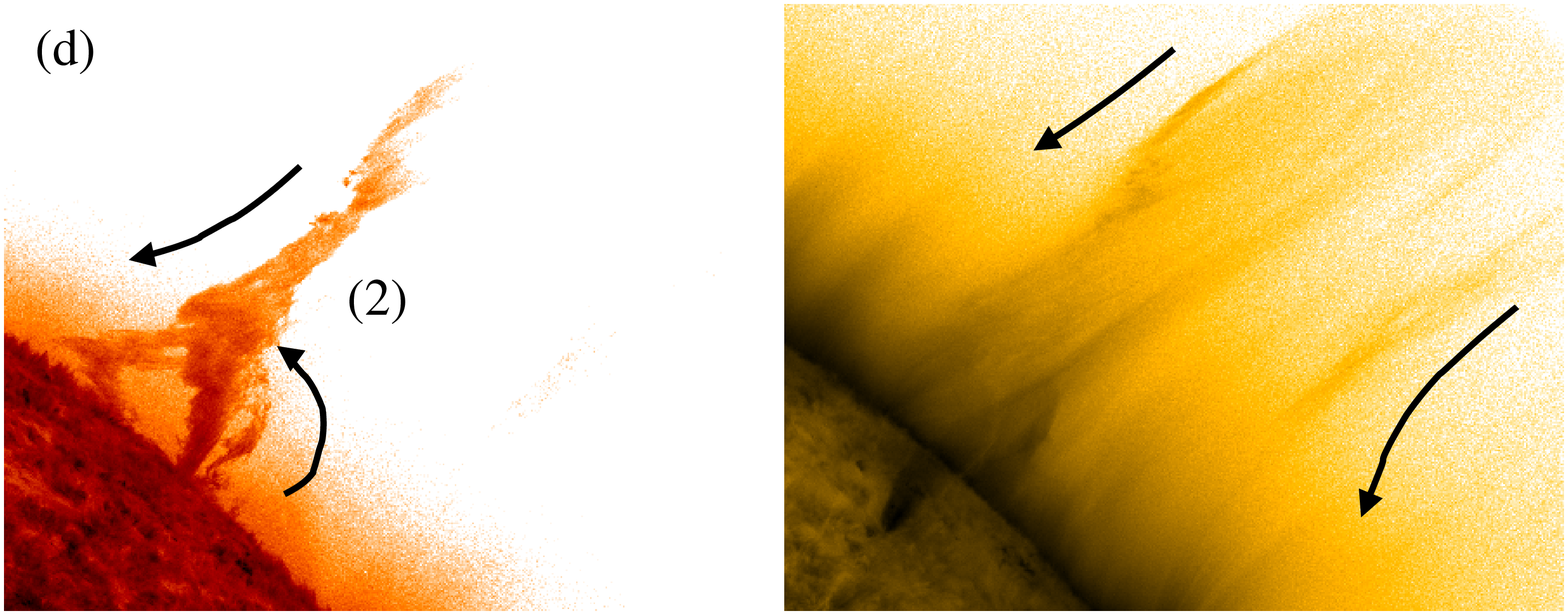}
\caption{Evolution of the polar crown cavity and the associated plasma (left) at
50000 K (304\AA) and (right) at 0.6 MK (171\AA) at three different times: (a)
00:03:12 UT, (b) 03:24:11 UT, (c) 06:51:11 UT and (d) 09:00:11 UT. The white
arrows indicate the direction of the plasma motions. The lines parallel to the
solar limb divide the different parts of plasma involved in the eruption (see
text).}
\label{fig:evol_cav}
\end{figure*}

Figs.~\ref{fig:closeup}a, b evidence the co-spatiality of the polar-crown material
in both the 304\AA~and 171\AA~channels . These snapshots highlight: 
\begin{itemize}
\item[-]{the
upward (U-shaped) bending of magnetic field lines at the bottom of the inner
cavity,}
\item[-]{the accumulation of prominence material along these field lines
suggesting that a magnetohydrostatic equilibrium is in place with the magnetic
field curvature acting against the gravity.}
\end{itemize}

The latter assumption is also supported by the fact that the cavity has been stable
for several hours before the eruption. Even if the location is similar, the
U-shaped field lines in both  channels are not filled by the corresponding
plasma in the same manner:  in the 171\AA~channel the length along the U-shaped
field lines filled by the coronal plasma appears longer than in the
304\AA~channel.

It is important to remember here that the observed structure is
integrated along the line-of-sight. Therefore limited three dimensional
depth information can be derived on the polar crown filament solely from this SDO/AIA
dataset only.

\section{Evolution of the cavity}
\label{sec:cavity}

In order to study the dynamics of the eruption, we first look at several
snapshots of the cavity to describe the motions and structures of the polar
crown filament. We restrict the dynamical study to the 304\AA~and
171\AA~channels in which the U-shaped structures are clearly seen and also for
which there is a
minimum of confusion with the background emission (see
Fig.~\ref{fig:closeup}c, d). 

Fig.~\ref{fig:evol_cav} outlines a series of images at four characteristic
times of the cavity evolution: (a) at 00:03 UT when the cavity is stable at a
height of 100 Mm above the surface (projection on the plane of the sky), (b) at
03:24 UT in the early phase of the eruption, (c) at 06:51 UT towards the end of
cavity eruption within the SDO/AIA field-of-view, (d) at 09:00 UT, a couple of
hours after the cavity has moved into the higher part of the corona, and the
plasma and magnetic field lines are still in the process of reorganisation and
relaxation. 

The polar crown material is divided into two parts, namely P1 and P2, that are
not distinguishable at first (Fig.~\ref{fig:evol_cav}a) but can be
differentiated in the following frames (Fig.~\ref{fig:evol_cav}b-c by the solid
line parallel to the limb). P1 corresponds to the main part of the eruptive
cavity. In Fig.~\ref{fig:evol_cav}b, the plasma contained in U-shaped fied lines
starts to move upwards (as indicated by the left arrow) whilst the plasma on the
right-hand side exhibits upwards flows along field lines (right arrow). P2
remains stable. In Fig.~\ref{fig:evol_cav}c, the plasma in P1 is detached and
thus ejected into the high corona. P2 starts to rise. 

In Fig.~\ref{fig:evol_cav}d, only the plasma in P2 remains at this height in the
corona whilst P1 continues its way out of the corona. The plasma in P2 is
flowing down towards the low corona following the field lines in both channels
(see arrows). From this time series, it is important to notice that the plasma
in both EUV channels are located at the same place below the polar crown cavity,
and this is only during the eruptive phase that the decoupling between the two
is observed. 

\begin{figure}[!t]
\centering
\includegraphics[width=1.\linewidth]{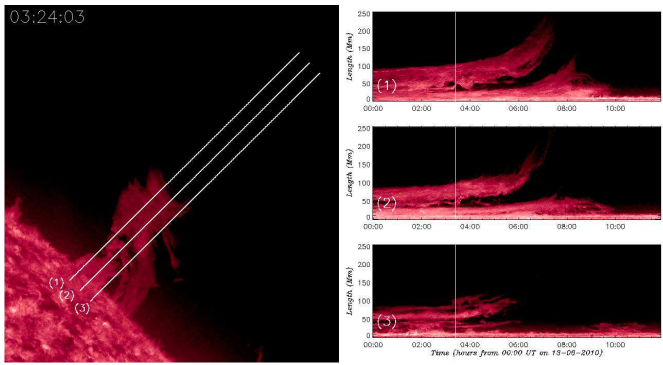}
\caption{Time slices of the eruptive cavity as observed in 304\AA~on 13 June
2010 by SDO/AIA. The image on the left has been taken at 03:24:12 UT in log
scale to highlight the cavity. On the right, the time slices are plotted along
three different radial directions: (1) and (3) are on the sides of the cavity,
(2) is at the minimum of the cavity. The three location are chosen at the
beginning of the time series running from 00:00 UT to 12:00 UT. The vertical
axis represents the distance in Mm from the bottom (on the disc) of the time
slices. The white vertical line indicates the time of the image on the left. The
movie is provided as online material.}
\label{fig:cut304}
\end{figure}

\begin{figure}[!t]
\centering
\includegraphics[width=1.\linewidth]{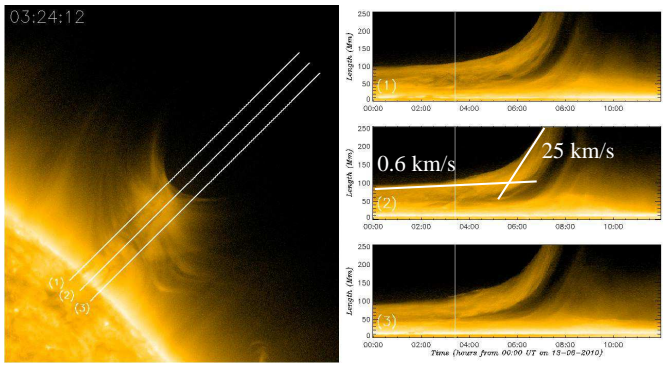}
\caption{Same as Fig.~\ref{fig:cut304} for the 171\AA~channel. The asymptotic
behaviour is indicated by the straight white lines. The estimated speeds are
also indicated. The movie is provided as online material.}
\label{fig:cut171}
\end{figure}

Second, we examine the radial evolution of the cavity at three different
locations along the width of the cavity by plotting three adjacent time slices
(Figs.~\ref{fig:cut304} and~\ref{fig:cut171}): the middle location (2)
corresponds to the minimum height of the cavity above the surface at the
beginning of the time series, whilst locations (1) and (3) are symmetrically on
both sides of the minimum height. The cavity appears in the top left corner of
the time slices and the bottom of the cavity is first located at about 90 Mm at
304\AA~and 100 Mm at 171\AA. Even if the motion of the cavity is not
entirely in the radial direction, the three time slices show the behaviour of a
large portion of the cavity during the eruption: the similar evolution in all
three slices supports the assumption that the cavity evolves as a solid body.
We note that the cavity leaves the field-of-view (250 Mm high) at about the same
time in both wavelengths (around 07:00 UT).

In Fig.~\ref{fig:cut304}, the time slices evidence the evolution of the cool
plasma at 304\AA~during the eruption process: the slow rise of the cavity during 4-5 hours
and then the eruption. The plasma of the cavity observed at 304\AA~does not
reach heights above 250 Mm. During the eruption process most of the plasma is
drained back down along magnetic structures as highlighted in
Fig.~\ref{fig:evol_cav}.  

From Fig.~\ref{fig:cut171}, we define the initial and final stages of the
eruption as observed at 171\AA~from the asymptotic behaviour of the time slice (2) as indicated by the
straight white lines. At the start, the cavity is stable justifying why we
started our study at 00:00 UT. The first phase is a slow rise of the cavity with
a characteristic speed of 0.6 km$\cdot$s$^{-1}$. The cavity follows this trend
until 03:00 UT. The second phase of the eruption is the faster motion of the
cavity with a characteristic speed of 25 km$\cdot$s$^{-1}$. Even if this speed
is far less than local Alfv\'en or sound speeds, this is comparable to the speed
of plasmoid ejection as reported by \citet{tsu97}.

\section{Discussion and Conclusions}
\label{sec:disc}

We propose to define a polar crown cavity as a density depletion at the bottom
of which the polar crown filament material sits indicating the existence of a
magnetohydrostatic equilibrium. The filament material is drained down along the
polar crown cavity by gravity and sustained by the action of the upward-directed
magnetic field curvature force. This fact as well as the long, steady
observations of the polar crown cavity indicate that the cavity is in a
magnetohydrostatic equilibrium. The cold and hot coronal plasma are located at a
similar location along the same field lines. The observations of the cavity
structure and plasma spatial distribution are consistent with the classical 2D
cartoon of a cavity depicted by concentric ellipses. For instance, in the
classical {\em  CSHKP} model \citep{car64,stu68,hir74,kop76}, the eruptive
structure is composed of a twisted flux tube at the bottom of which the plasma
is concentrated. Contrary to the cartoon proposed by \citet{cli86} placing the
filament material at the centre of the cavity, the filament material is located
at the bottom of the cavity which is more consistent with the model of
\citet{mar89}.

In the observations reported here,  magnetic curvature compensates gravity to
create an equilibrium state in which the density is considerably increased at
the bottom of the cavity. We also show that the flows along field lines and
varying from one wavelength to an other are important for the initiation
(Fig.~\ref{fig:evol_cav}b) and relaxation (Fig.~\ref{fig:evol_cav}d) phases of the
eruption. We also note that the rise of the cavity (divided in two stages) is
similar to the  plasmoid eruption initiated by an impulsive flare as reported by
\citet{ohy97} \citep[see also][]{shi98}.  

This preliminary study describes the structure and evolution of a polar crown
filament and its cavity projected onto the plane of the sky and, in any case,
gives a full picture of the eruption process. In a forthcoming paper, we will
discuss the possible triggers of the cavity eruption of concomitant external
(flares, CMEs) and internal (kink/torus instability, mass loading) phenomena by
combining SDO/AIA and STEREO/SECCHI/EUVI images which give us a more realistic
3D representation of the event. 

\begin{acknowledgements}
We thank the referee for his/her useful comments helping to improve the letter.
The SDO/AIA data have been obtained through the University of Central Lancashire
database. The data used are provided courtesy of NASA/SDO and the AIA science
team. CEA acknowledges the support of the STFC studentship programme. The
SOHO/LASCO CME catalog is generated and maintained at the CDAW Data Center by
NASA and The Catholic University of America in cooperation with the Naval
Research Laboratory. SOHO is a project of international cooperation between ESA
and NASA.
\end{acknowledgements}

\bibliographystyle{aa}




\bibliography{/wisdom/work/sr1/TEX/Bib/mybib}

\Online

\begin{appendix} 
\section{Time evolution of the cavity}

Movies of the evolution of the cavity during the eruption and the associated
time slices in both 304\AA~and 171\AA~channels (see Figs.~\ref{fig:cut304}
and~\ref{fig:cut171}) are supplied as online material.

\end{appendix}

\end{document}